# Frequency fluctuations in silicon nanoresonators

Marc Sansa[1,2], Eric Sage[1,2], Elizabeth C. Bullard[3], Marc Gély[1,2], Thomas Alava[1,2], Eric Colinet[1,2,†], Akshay K. Naik[4], Luis Guillermo Villanueva[5], Laurent Duraffourg[1,2], Michael L. Roukes[3], Guillaume Jourdan[1,2], Sébastien Hentz[1,2*]

1. Univ. Grenoble Alpes, F-38000 Grenoble, France
2. CEA, LETI, Minatec Campus, F-38054 Grenoble, France
3. Kavli Nanoscience Institute and Departments of Physics, Applied Physics, and Bioengineering, California Institute of Technology, MC 149-33, Pasadena, California 91125 USA
4. Centre for Nano Science and Engineering, Indian Institute of Science, Bangalore, 560012, India
5. Advanced NEMS Group, École Polytechnique Fédérale de Lausanne (EPFL), CH-1015 Lausanne, Switzerland

† Present address: APIX Analytics, 7 parvis Louis Néel – CS20050, 38040 Grenoble cedex 09, France

* E-mail : sebastien.hentz@cea.fr

*Frequency stability is key to performance of nanoresonators and their applications. This stability is thought to reach a limit with the resonator's ability to resolve thermally-induced vibrations. Although measurements and predictions of resonator stability usually disregard fluctuations in the mechanical frequency response, these fluctuations have recently attracted considerable theoretical interest. However, their existence is very difficult to demonstrate experimentally. Here, through a literature review, we show that all studies of frequency stability report values several orders of magnitude larger than the limit imposed by thermomechanical noise. We studied a monocrystalline silicon nanoresonator at room temperature, and found a similar discrepancy despite the ability to resolve thermal noise. We propose a new method to show this was due to the presence of frequency fluctuations, of unexpected level. The fluctuations were not due to the instrumentation system, or to any other of the known sources investigated. These results challenge our current understanding of frequency fluctuations and call for a change in practices.*

Nano-electro-mechanical systems (NEMS) have demonstrated their tremendous potential for both basic science and industrial applications. These systems have opened a new window into the realm of quantum physics[1,2] and non-linear dynamics[3,4] and allow record limits of detection in high-performance force[5] and mass[6] sensing. These records have been achieved through extreme miniaturization, thanks to the advent of carbon nanotubes (CNTs) and graphene monolayer sheets. Indeed, the minimum mass (or force) detectable by a resonator is proportional to its total mass (or stiffness). This limit-of-detection is also proportional to the measurement uncertainty of the resonance frequency, $<\frac{\delta f}{f_0}>$, therefore much work has been dedicated to determining the limits of the frequency stability of nanomechanical resonators[7,8].

Frequency stability can be affected by noise added to the signal amplitude, provoking jitter in the phase (hereafter *additive phase noise*) or by fluctuations in the device's overall mechanical response, inducing spectral broadening and resonance frequency fluctuations (hereafter *frequency fluctuations*)[9].

The frequency stability and limit-of-detection for a device are commonly predicted based on the dynamic range (DR) measured[10–12] (ratio between maximum driven signal level and noise floor expressed in dB) by applying the simple formula[13,14], $\langle \frac{\delta f}{f_0} \rangle = \frac{1}{2Q} 10^{-\frac{DR}{20}}$. Additive phase noise generally comes from the device being coupled to a thermal bath. The DR formula implies that, for a given drive level, frequency stability is maximized when the random motion of a resonator driven by thermomechanical noise can be resolved, which has led to considerable efforts over the past decade to design nanoscale systems in which transduction is efficient[5,15,16]. However, the formula holds true in conditions where frequency fluctuations can be neglected, which is almost never verified, partly because it is not trivial to distinguish additive phase noise from frequency fluctuations[17–19]. Nevertheless, numerous sources of frequency fluctuations have been theoretically described, including adsorption-desorption noise[7,8,20], temperature noise due to finite heat capacity[8], defect motion[7] or molecule diffusion along the resonator[9]. Although this issue has attracted considerable theoretical interest, very few experimental studies have observed the signature of one or more of these sources of fluctuations[21,22]. Instead, fluctuations in device temperature, in charge state or in stiffness due to signals in the instrumentation are thought to explain most observations of frequency fluctuations[18,23–25]. Moreover, these observations were only possible at low temperature with devices particularly susceptible to fluctuations like ultra-high Q devices[22] or CNTs[18,24] and graphene membranes[25].

We begin this article with a comprehensive review of published frequency stability studies. This review reveals that the limit imposed by thermomechanical noise has never been reached across a wide range of devices, and that the experimentally observed frequency stability values exceed the thermomechanical noise limit by several orders of magnitude. To better understand this phenomenon, we tested a canonical, CMOS-compatible monocrystalline silicon nanoresonator and found a discrepancy of similar magnitude at room temperature, even though thermally-induced vibrations were well-resolved. Analysis of the correlation properties of the excess noise showed that the mechanical frequency response fluctuates as a whole. Thus, as it ignores frequency fluctuations, the well-established DR formula falls several orders of magnitude short when used to predict the frequency stability of these devices. We also found that frequency fluctuations are not due to the instrumentation, nor to a range of known sources.

These results call for further investigation of the microscopic mechanisms causing frequency fluctuations, which had not been observed in semiconductor-grade silicon resonators and oscillators. In light of these findings, many past experiments and predictions of frequency stability or limit-of-detection made based on the DR formula, which only considers additive phase noise, must be revisited.

**LITERATURE REVIEW**

In this work, the frequency stability $<\delta f/f_0>$ was estimated with the Allan deviation $\sigma_A$ (see Methods) [26]. This metrology standard is particularly suited to practical integration times and is complementary to power spectral density measurements in the frequency domain. In Figure 1 we plot the Allan deviation of published results that provide measurements for the frequency stability against the total mass of the different devices studied. We have tried to be exhaustive in our review of stability studies on nanoscale resonators. The articles reviewed encompass a large range of dimensions (over 15 orders of magnitude in device mass) and technologies: flexural-mode micro-resonators (MEMS), top-down nanoresonators (NEMS), and bottom-up nanoresonators (CNTs and graphene devices). The reported frequency stabilities are compared with the limit imposed by the theoretical thermomechanical noise, estimated with the DR formula. To improve this comparison, a normalization factor for temperature and pressure was applied across studies (see Supplementary Section 1).

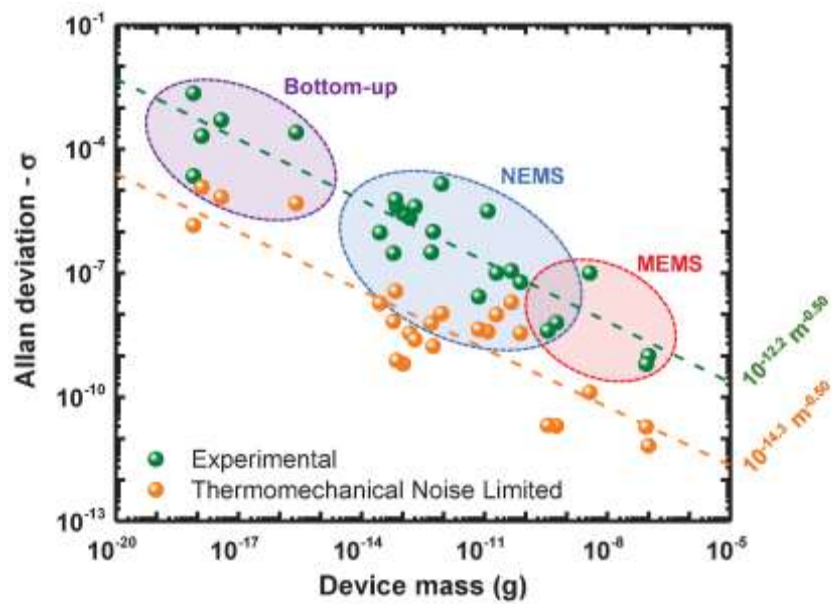

**Figure 1. The frequency stability of resonators measured in the literature is on average 2.1 orders of magnitude greater than the thermomechanical-noise-limited stability.** For each device, both the experimentally measured frequency stability (green) and the

analytically calculated thermomechanical limit at a temperature of 300 K for the frequency determination (orange) are plotted. The dependence of both magnitudes with the mass of the device is similar $\sim m^{-1/2}$. The dashed lines represent the best fit for each set of data (thermomechanical-noise-limited and experimental). Supplementary Figure S1 shows complete mapping of the references with the datapoints.

Despite the considerable experimental variety, Figure 1 shows a very clear picture: none of the studies reviewed attained the frequency stability limit set by thermomechanical noise. The experimental results were always at least an order of magnitude greater than the theoretical limit, and on average 2.1 orders of magnitude greater (the same conclusions can be drawn from the non-normalized data, see supplementary Figure S2). Interestingly, this observation holds true from MEMS to CNT resonators, even though dynamic range decreases with device size[27]; the best linear fits of both experimental stability and thermomechanical limit scale similarly for all device types at $\sim m^{-1/2}$. The discrepancy has been noted across a large variety of designs and resonating modes: of the 25 datapoints, 6 correspond to flexural mode in clamped-free beams[16,28–32], 15 correspond to flexural mode in clamped-clamped beams (3 of which were tensile stressed)[6,11,22,33–43], 2 correspond to flexural mode in pinned beams[35,44], and 2 correspond to flexural mode in thin membranes[45,46]. Similarly, no differences due to transduction techniques, optical detection[22,29,30,32,42,43], capacitive[40,41,46], magnetomotive[36–38], piezoelectric[31,44], piezoresistive[16,34,35,39] or field-effect-modulated conductance[6,11,28,33,45] were observed. The limiting factor in frequency stability was seldom discussed; in two cases[31,41], the Signal-to-Noise Ratio (SNR) was limited by the amplifier noise and in some others, the authors suggest that extrinsic sources of frequency fluctuations - like noise in the drive signal or temperature fluctuations[39,44] - may dominate. Nevertheless, it remains intriguing that, despite the great effort expended to do so (particularly in the "NEMS" sub-group), the thermomechanical noise limit was never reached in any case. This huge discrepancy was never discussed, and nor was the validity of the DR formula. We believe that further exploration of the issue is warranted, and we provide it in this article with a simple device made from a high-quality material.

**FREQUENCY STABILITY IN MONOCRYSTALLINE Si RESONATORS**

To follow-up on the conclusions from the literature review, a series of experiments was performed on monocrystalline silicon resonators fabricated from Silicon-On-Insulator wafers with Very Large Scale Integration processes[16], at room temperature (unless otherwise stated) and typical pressure of $10^{-5}$ Torr. The resonators were electrostatically actuated and use a differential piezoresistive readout (see Figure 2a). The downmixing set-up used was sensitive enough to measure the thermomechanical noise of the resonator, which was 2.5 times larger than our experimental noise floor (Figure 2b), and yielded a very large linear dynamic range (~107 dB for 1 s integration time, see Supplementary Figure S3). These features make these resonators high-performance gravimetric sensors[47]. Fabrication and measurement details can be found in Methods and in Supplementary Sections 2 and 3.

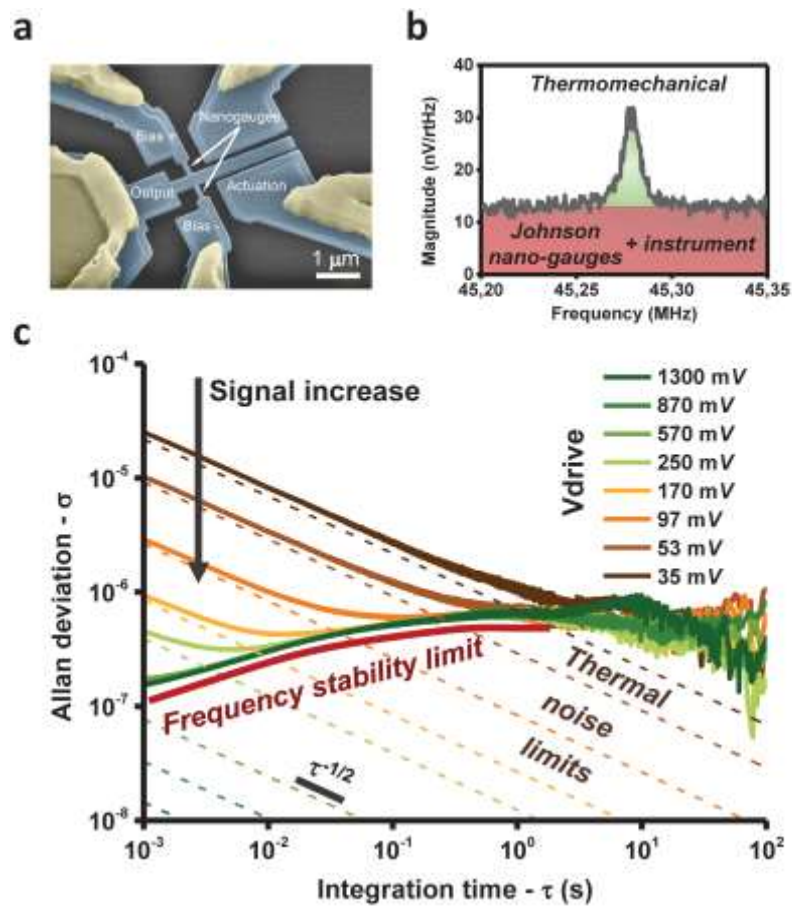

**Figure 2. The frequency stability of our monocrystalline silicon nanomechanical resonators is limited by a source of noise exceeding thermal fluctuations**. **a,** Crystalline Si NEMS resonator used to perform measurements. Typical dimensions are 3.2 µm (length), 300 nm (width), 160 nm (thickness). The piezoresistive nanogauges are typically 1 µm long and 100 nm wide. **b,** Spectrum of the thermomechanical noise measured in the resonators studied. The noise floor was determined from Johnson noise in the nanogauges and contacts, and noise from the readout instrumentation (lock-in amplifier). Typical quality factors were 5000-7000 at room temperature. **c,** Allan deviation as a function of integration time, from 1 ms to 100 s. This range was chosen as the response time of the resonator was $\sim \frac{2Q}{f_0} = 0.25$ ms, with a readout instrumentation limit of 50 µs, and because systematic drifts occur after ~100 s (see Supplementary Figure S4). Drive voltage amplitudes were chosen from 35 mV (yielding a Signal-to-Noise Ratio (SNR) of about 62.5 for a measurement bandwidth of 1 Hz)

to 1.3 V (yielding a displacement of about half the onset of non-linearity, see Supplementary Figure S3). The bias voltage amplitude was maintained constant at 1.5 V. The dashed lines indicate the expected stability from the output signal at each drive voltage and the total additive noise in the system, as measured in panel b), see equation (1). The red line is a visual guide, highlighting the experimentally measured lower bound for frequency stability. This bound is several orders of magnitude higher than the expected one.

The resonance frequency of the resonator was deduced from its open-loop phase fluctuations (see Methods). The resulting experimental Allan deviation, $\sigma_A$, is illustrated by the solid lines in Figure 2c, for integration times covering five orders of magnitude.

The dashed lines in Figure 2c show the theoretical Allan deviation, which would be expected in a regime where additive phase noise dominates the frequency stability, based on the DR formula[14] expressed in the voltage domain:

$$\sigma_A \cong \frac{1}{2Q}\frac{N_T}{S}\sqrt{\frac{1}{2\pi\tau}} \qquad (1)$$

where Q is the quality factor of the resonator (see details in Supplementary Section 4), S is the amplitude of the output signal at the resonance frequency for each drive (in V, see Supplementary Figure S3), $N_T$ is the noise level at the output (32 $nV\ Hz^{-1/2}$ in our case), $\tau$ the integration time ($1/2\pi\tau$ is the measurement bandwidth with a first-order low-pass filter). The SNR for the measurement is therefore $\frac{N_T}{S}\sqrt{\frac{1}{2\pi\tau}}$ (equal to phase fluctuations, see Supplementary Figure S5). For a dominant additive white noise, the expected Allan deviation scales like $\tau^{-1/2}$, and is inversely proportional to the output signal, S.

Figure 2c clearly shows that equation (1) accurately describes the frequency stability of our resonators for short integration times and low drive amplitudes. This result suggests that within this range, the system is in a regime where additive phase noise dominates frequency stability. However, at higher drive amplitudes and for longer integration times, the experimental observation significantly deviates from the expected behavior. The red line in Figure 2c indicates the lower bound for resonator frequency stability, which cannot be improved below this limit by increasing the drive amplitude. The Allan deviation first increases and subsequently varies little with integration time. This latter behavior is consistent with plots of power spectral density (see Supplementary Figure S6), where the major trend appears to be a slope of 1/f for high drive. As a result, the limit-of-detection for this NEMS is more than two orders of magnitude higher than expected for a typical measurement time of 100 ms. These results are consistent with the presence of frequency fluctuations (see also in-phase and quadrature plots in Supplementary Figure S7). Nevertheless, these fluctuations were quite unexpected for devices made from a high-quality material like monocrystalline silicon. Moreover, the level of the discrepancy – several orders of magnitude – is even more surprising given that the measurements were performed at room temperature in relatively straightforward experimental conditions. A similar discrepancy was observed in all our experimental set-ups, regardless of location, as well as with clamped-clamped beam resonators fabricated using the same technology (see Supplementary Figure S8).

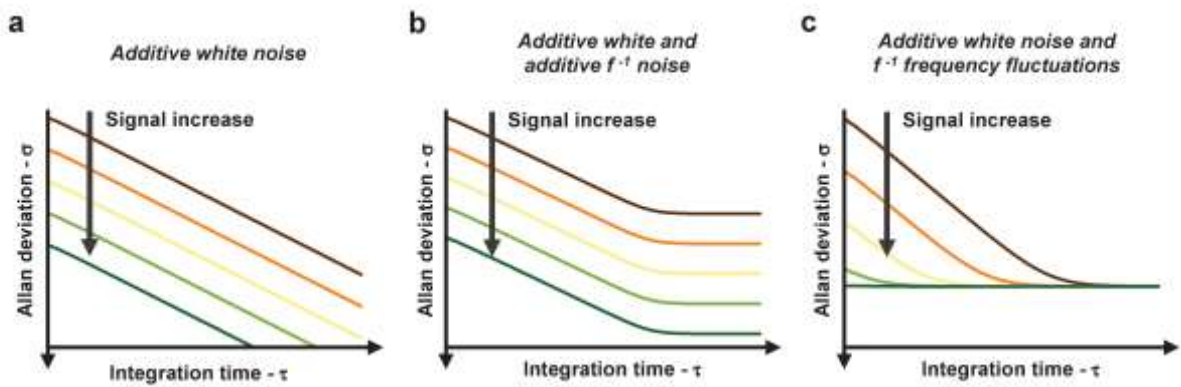

**Figure 3. Additive phase noise and frequency fluctuations show different features in the Allan deviation.** Effect of different noise sources on the frequency stability as a function of the integration time $\tau$, and for different signal levels. **a,** Additive white noise, manifesting itself as phase noise. It presents a constant slope of $\tau^{-1/2}$. The stability improves with increasing signal level. **b,** Combination of additive white and $f^{-1}$ noises. For low integration times it presents a slope of $\tau^{-1/2}$, which becomes $\tau^0$ when the $f^{-1}$ noise dominates at large integration times. The stability improves with increasing signal level in the whole time range. **c,** Combination of additive white noise with $f^{-1}$ frequency-fluctuations. For low integration times it presents a slope of $\tau^{-1/2}$, which becomes $\tau^0$ when the $f^{-1}$ frequency noise dominates. Moreover, the stability due to frequency fluctuations is insensitive to the signal level: therefore, an increase in the signal has an effect only when additive noise dominates.

### NATURE OF THE EXCESS NOISE IN SILICON RESONATORS

The lower bound for the Allan deviation (red line in Figure 2c) does not depend on drive level. This would be the case in the presence of a source of frequency fluctuations $N_f$ which would add to the additive noise-limited stability in equation (1): $(<\delta f> \approx \frac{f_0}{2Q}\frac{N_T}{S}\sqrt{\frac{1}{2\pi\tau}} + N_f)$. It would also be the case if the additive noise was proportional to signal amplitude ($N_T \propto S$ in equation (1)). This is illustrated in Figure 3 and Supplementary Section 4. The presence of non-linear damping could also limit the improvement of frequency stability with increasing drive, but our devices do not display any significant non-linear damping (see Supplementary Figure S3). It should be noted that spectral broadening is not observed with our devices either: ring-down measurements give the same linewidth as the spectral measurements (see Supplementary Figure S17).

White noise probed simultaneously at two different frequencies is uncorrelated[14], conversely, frequency fluctuations induce a shift in the whole frequency response of the resonator ; thus, probing noise at two different frequencies within the resonator's bandwidth should show strong correlation in the case of dominant frequency fluctuations (see Figure 4a). The correlation properties of the observed noise were therefore studied as a function of integration time and drive amplitude.

Two distinct frequency traces were simultaneously recorded, and their stability was assessed by plotting their Allan deviation (Figure 4a, see Methods and Supplementary Figure S9). The result (Figure 4b) was very consistent with the results shown in Figure 2c, and was almost identical for the two frequency traces (Supplementary Figure S10). We computed the correlation of the pair of frequency traces (see Methods) from this data set (Figure 4b).

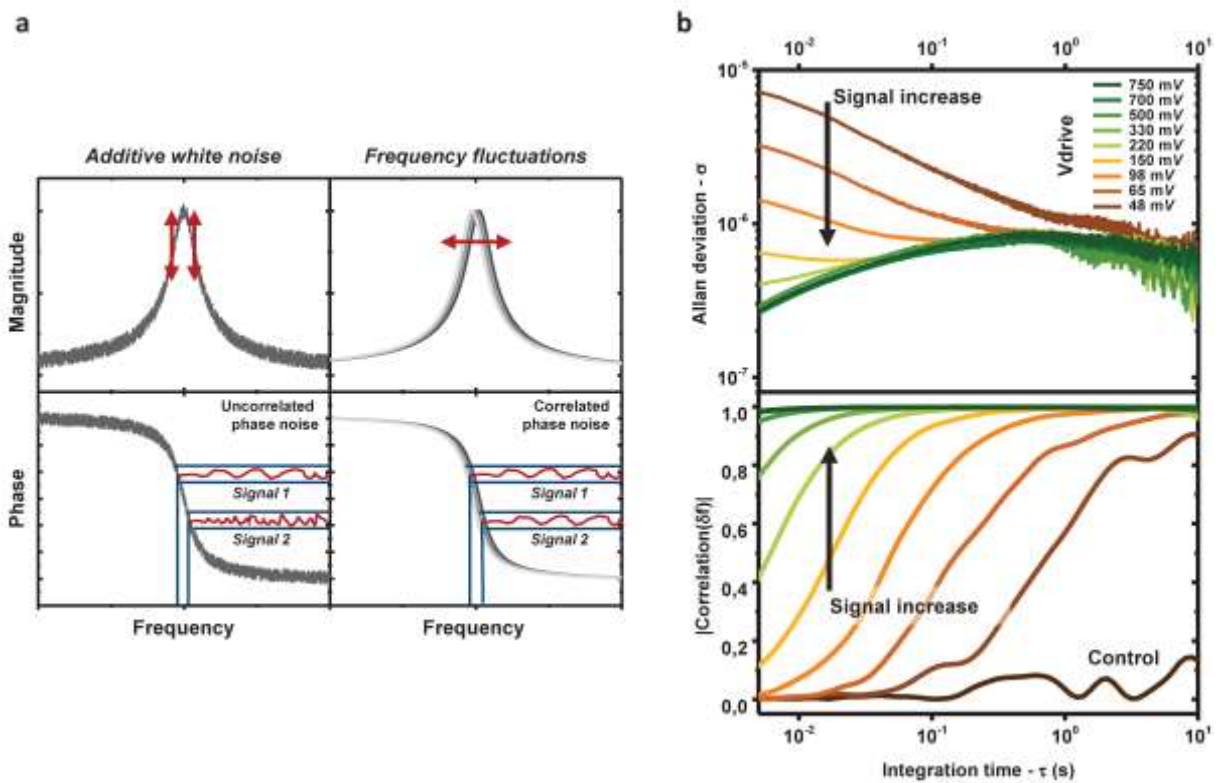

**Figure 4. The limit in frequency stability of our silicon resonators is due to frequency fluctuations. a,** The resonator was actuated at two different frequencies within its bandwidth, typically at ±1 kHz from the central resonance frequency. The stability of each independently-obtained frequency trace was estimated from the open-loop phase information ($f(t) \propto \phi(t)$ for small deviations from the resonance frequency). An additive white noise source is uncorrelated at different frequencies. Response signals measured at different frequencies within the bandwidth are then also uncorrelated. In contrast, frequency fluctuations shift the whole frequency response of the resonator. Response signals measured at different frequencies are then strongly correlated. **b,** (top) Allan deviation of one of the frequency traces obtained using this measurement method. The other trace presents very similar stability results (Supplementary Figure S10). The results are consistent with the single-frequency measurements shown in Figure 2c. (bottom) Correlation between the two simultaneous frequency traces for the same sample set. As expected, the correlation was weak when the noise was dominated by additive phase noise (low drive amplitudes), but the correlation was high at long integration times. This time range depends on the drive level. The "control" curve shows the same experiment performed out of resonance, at maximum drive voltage. These results indicate the existence of fluctuations of the whole frequency response of the resonator, *i.e.*, frequency fluctuations.

The correlation is thus closely linked to the integration time and the drive voltage; Figure 4 clearly indicates that the signals are weakly correlated when the dominant noise is additive white noise (low drive levels), and strongly correlated when the excess noise is dominant (*i.e.*, at long integration times for low drive levels or over the whole time range for high-enough drive levels). Control measurements were also taken, choosing the two sideband frequencies out of resonance (but maintaining a constant difference). In these conditions, no correlation was observed whatever the drive voltage (Figure 4b and Supplementary Figure S11). The only difference between this control and the in-resonance measurements was the almost total absence of mechanical response in the control. This result indicates that the limit in frequency stability observed with our silicon nanomechanical resonators is due to fluctuations of the resonator's overall frequency response in the mechanical domain, *i.e.* frequency fluctuations (as opposed to some type of noise in the measurement system downstream of the piezoresistive transduction).

DISCUSSION OF THE PHYSICAL ORIGIN OF THE FREQUENCY FLUCTUATIONS

In the vast majority of studies where frequency fluctuations were thought to explain experimental observations, the source of these fluctuations was noise due to the instrumentation[18,24,25,39,44,48]. In this study, we started by eliminating sources of noise present in the instrumentation, such as the frequency stability of the drive signal. Amplitude noise in this signal also leads to frequency shifts due to the non-linear Duffing term in the equation of motion, or due to electrostatically-induced changes in stiffness. Similarly, bias signal shifts frequency because of Joule heating. In our system, experimental characterization of these sources of frequency fluctuations showed that none of them could explain our observations (see Supplementary Figures S12 and S13).

Variations in device temperature can also lead to frequency fluctuations, with a typical temperature coefficient of $-50\ ppm\ K^{-1}$. However, these fluctuations can be compensated for by using the second mode frequency as a temperature probe. In our experiments, we tracked frequency fluctuations of two modes and used the frequency fluctuations of one of these modes to correct for temperature-induced variations on the other. This correction did not significantly improve the Allan deviation (Figure 5 and Supplementary Section 5).

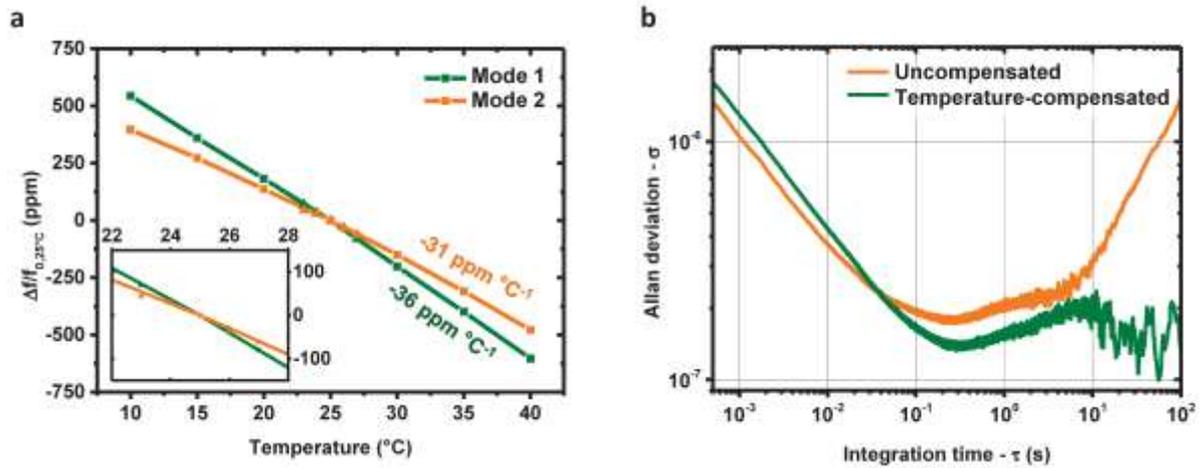

**Figure 5. The limiting frequency fluctuations are not due to temperature fluctuations alone. a,** Temperature dependence of the first two modes of the resonator, obtained by measuring their resonance frequency for a range of temperatures around 25 °C. The squares represent a coarse measurement for a wide range of temperatures, the triangles a detailed measurement around room temperature (-38.2 $ppm\ °C^{-1}$, R-square error 0.999 for the first mode; -29.1 $ppm\ °C^{-1}$, R-square error 0.997 for the second mode). The inset shows a detail of the sensitivity around room temperature (-36.4 $ppm\ °C^{-1}$, R-square error 0.993 for the first mode; -27.6 $ppm\ °C^{-1}$, R-square error 0.982 for the second mode). **b,** Frequency stability of the first mode before (orange) and after (green) temperature correction. Three regimes are clearly visible on this plot: In the white noise regime ($\tau < 10^{-1}$ s), temperature compensation slightly degrades frequency stability, as it is the addition of uncorrelated white noise of both modes ($10^{-6}$ and $6.5 \times 10^{-7}$ for $\tau$=1 ms, quadratically summing to $1.2 \times 10^{-6}$; the temperature compensated deviation is found to be $1.25 \times 10^{-6}$). With integration times of $\tau > 10^1$ s, long-term drifts can be measured, in this region, stability was improved by compensation for temperature-induced drifts in resonance frequency. In the frequency fluctuations regime ($10^{-1}\ s < \tau < 10^1\ s$), no significant improvement was observed.

Frequency fluctuations are also often attributed to molecules randomly adsorbing and desorbing onto/from the resonator, or diffusing along its surface. Models for these two sources exist and have been confronted to experiments in past studies[21] (see Supplementary Section 6). Frequency fluctuations can also be caused by thermalization of higher-order modes through non-linear mode coupling[25,49–52]: the frequency of one particular mode depends on the vibration amplitude of the other modes because of stiffness-induced coupling (a particular case is the dependence of one mode frequency on the amplitude of motion of this mode via the Duffing term). The contributions of modes 1 and 2 are dominant in these coupling effects in our case (see Supplementary Section 6). We therefore measured the amplitude-to-frequency relationships of the resonator's first two modes and assumed thermally-induced vibrations to assess the coupling effects. This analysis is summarized in Figure 6, showing the Allan deviation induced by the sources discussed above. Although it would be useful to further investigate the mode coupling effect by studying the interrelation between the coupling and the decay rate of the contributing modes[53], our approach shows that each of the known sources tested, as well as the sum of all sources, is several orders of magnitude lower than the overall experimental frequency instability.

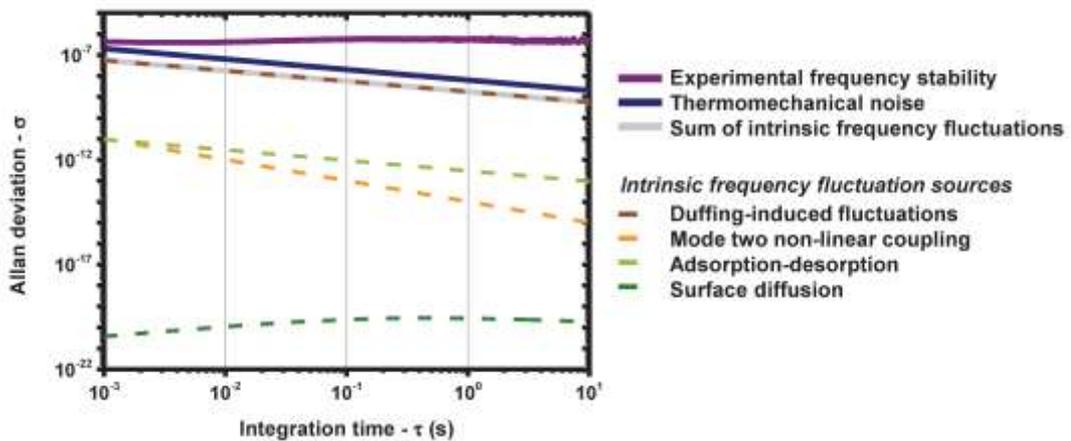

**Figure 6. Known sources of frequency fluctuations.** The frequency fluctuations caused by different sources of noise, and comparison with the thermomechanical noise limit (thick blue line) and experimental results (thick violet line) were estimated in a clamped-clamped beam resonator. Frequency fluctuations arising from adsorption-desorption and surface diffusion were calculated using theoretical models. Thermomechanical noise is also a source of frequency fluctuations, through Duffing non-linearity. The coupling between the amplitude of motion of mode 2 and the resonance frequency of mode 1 was experimentally characterized, and the thermomechanical noise-induced vibrations of mode 2 are measured to quantify the resulting frequency fluctuations. The thick gray line indicates the sum of fluctuations due to these four sources of frequency fluctuations. This level of fluctuation is lower than the thermomechanical noise limit, and orders of magnitude lower than the experimental frequency instability.

Few known mechanisms remain to be explored. Bulk and surface effects are likely to play an important role in the frequency fluctuations observed. Dielectric- and charge fluctuations have been reported to cause frequency fluctuations in various microscopy probes due to interaction with nearby surfaces (at a few tens nm distance)[23,54]. In the case of our nanoresonator, charges can move on and off traps present at the surface of the silicon due to native oxide formation. This charge motion will induce frequency fluctuations through electrostatic stiffness. The magnitude of frequency fluctuations due to charge fluctuators is expected to vary considerably with the actuation gap (to the power of 3) and with drive voltage[23]. However, we observed no measurable change with these parameters. Furthermore, unlike in highly stressed amorphous silicon nitride resonators[22], the number of defects in the bulk of pure monocrystalline silicon nanoresonators is too low to provide a significant source of frequency fluctuations due to defect motion[7]. Nevertheless, two-level systems-like behavior could still be encountered due to, for example, the doping levels used.

## Conclusion

Frequency fluctuations have recently become a topic of considerable interest, mostly in basic research. These fluctuations are usually ignored in experiments aiming to assess nanoresonator performance or in the numerous cases where the DR formula is used to predict performance. A careful review of most published frequency stability measurements for nanoresonators showed that none of them attained the limit set by thermomechanical noise, and that the Allan deviation measured was on average more than two orders of magnitude higher than this limit. We investigated this point with a monocrystalline silicon nanoresonator and found a discrepancy of similar magnitude, even though random motion due to thermomechanical noise was well-resolved in the absence of coherent drive. Study of the correlation properties of the excess noise indicated that the whole mechanical frequency response fluctuated. We also found that these frequency fluctuations were not due to the instrumentation, but rather that they originated in the mechanical domain of the device. Fluctuations were not due to temperature variations, or to a range of other known sources such as adsorption-desorption noise. These results call for further investigation of microscopic mechanisms that could induce such frequency fluctuations, which had not previously been observed in semiconductor-grade silicon devices. The measured magnitude of these fluctuations is all the more unexpected, in particular at ambient temperature and in the absence of complex experimental conditions. These results suggest that we need to rethink a number of accepted assumptions, and make changes to current practices:

It is always assumed that increasing signal or decreasing additive phase noise (by, for example, improving transduction efficiency) improves frequency stability. This is not true in the presence of frequency fluctuations. Given the variety of devices used throughout the literature, it is possible that different mechanisms explain the limit found with different devices (Figure 1). However, it is not unlikely that frequency fluctuations, whatever their physical origin, are ubiquitous and are a major performance limiter for many nanoresonators. To confirm this paradigm shift, we believe many past and future experiments should be examined in light of our findings; many frequency stability predictions should also be reviewed because they applied the DR formula which omits frequency fluctuations. For example, the following methodology could be followed: the additive noise floor of the system should first be assessed by measuring the output signal of the undriven device (Figure 2b). The expected Allan deviation can be computed from this measurement for given drive levels. The corresponding experimental Allan deviations can be measured by recording the phase signal while driving the device at its resonance frequency. Plotting the Allan deviation is both simple and powerful to identify frequency fluctuations. These fluctuations can be further confirmed by the correlation technique proposed in this paper, which is a straightforward means to identify the presence of frequency fluctuations. Moreover, like the Allan deviation, it provides information on the temporal dynamics of these fluctuations at practical time scales. Finally the contribution of instrumentation to these fluctuations should be assessed to examine the physical mechanisms behind fluctuations originating in the mechanical domain of the device.

A great deal of modern technology relies on the purity of semiconductor electronics-grade silicon. For this reason, it is considered to have one of the highest mechanical qualities and it has thus recently become a commonly used material for commercial M/NEMS. Although significant experimental work remains to be done to elucidate the microscopic origin of the frequency fluctuations observed, our findings are of paramount importance for applications of a wide range of nano- (and possibly micro-) resonators, even those made of high-quality materials. Resonant mass (e.g. traces of low-mass volatile compounds), force (e.g. for Scanning Near-field Optical Microscopy or Magnetic Resonance Force Microscopy[55]) or inertial sensing, as well as time-reference devices, will no doubt benefit from further work on this topic.

**METHODS**

**Measurement of the frequency response and frequency stability.** The frequency response of the resonators was measured using a downmixing method, described in detail in [16]. The device was electrostatically actuated, and the driving voltage was applied to a side-gate parallel to the resonator. To reduce parasitic signals, the drive signal was set to half the actuation frequency $\frac{\omega}{2}$, thus the amplitude of motion of the resonator was proportional to the square of the actuation voltage. Motion of the resonator was detected differentially by two piezoresistive nanogauges. A bias voltage at $(\omega + \Delta\omega)$ through the gauges was used to down-mix their resistance change (occurring at the actuation frequency $\omega$), and the low-frequency readout signal at $\Delta\omega$ was detected using a lock-in amplifier. Typical measurement values were 1.5 V for the bias voltage at a measurement frequency of 500 kHz. All measurements were performed in a vacuum chamber at a pressure of $10^{-5}$ mbar and at room temperature. Thermomechanical noise was measured using the same set-up, with the drive electrode disconnected. Measurements were taken with a lock-in amplifier, which also generated the drive and bias signals.

The Allan deviation was measured in open-loop configuration, and the frequency stability was extracted from the response of the resonator actuated at resonance frequency with a fixed driving frequency. The phase of the measured signal, $\emptyset(t)$, was monitored for a certain amount of time, and then transformed into frequency fluctuations using the phase response of the resonator. Close to the resonance frequency, this phase response was linear, $\frac{\Delta\emptyset}{\Delta f} \cong \frac{2Q}{f_0}$. Using the complete phase response of the resonator instead of this linearization does not significantly alter the Allan deviation. Harmonics appearing at the frequency of the electricity supply (multiples of 50 Hz) were filtered out of data during post-processing.

Using this method, we obtained $N$ samples of the resonance frequency of the resonator $\overline{f_1} \cdots \overline{f_N}$, each averaged over an integration time, $\tau_0$. The Allan deviation for this integration time could then be defined as [26]:

$$\sigma_A(\tau_0) = \sqrt{\frac{1}{2(N-1)} \sum_{1}^{N-1} \left( \frac{\overline{f_{i+1}} - \overline{f_i}}{f_0} \right)^2} \qquad (2)$$

To obtain the frequency stability for higher integration times from the same set of frequency samples, we followed the standard method[26]. Initial samples were averaged in groups of *n* samples, and the Allan deviation for the new array was calculated using equation (2) to determine $\sigma_A(n\tau_0)$. This process was repeated multiple times until the number of samples was too low to provide a statistically significant result.

**Correlation measurements.** Correlation measurements were performed by simultaneously measuring the response of the resonator at different frequencies within the resonator's bandwidth. The measurement set-up was based on the one described in Supplementary Section 3, but here each signal was doubled, using two drive signals at different frequencies, two bias signals, and two measurement signals (Supplementary Figure S9 shows a detailed measurement scheme). Particular care was taken when choosing the drive signal amplitudes so that the resonator remained in the linear regime. Moreover, the two measurement frequencies were chosen to avoid cross-talk (e.g. 302 kHz and 367 kHz). Measurements were taken with the same lock-in amplifier input to ensure simultaneity. Although here we used a down-mixing set-up, correlation could also be measured with a homodyne method.

The phase traces were converted to frequency traces corresponding to the different integration times, as described above. Here, the complete phase response of the resonator was used rather than the linear approximation, as the frequencies for phase samples can be quite different from the resonance frequency. With this method we obtained two frequency sample arrays with an integration time $\tau_0$.

The graph in Figure 4b shows the correlation of these frequency traces versus the integration time $\tau$. We processed the signals so that the correlation for a given $\tau$ only depends on frequency variations with characteristic time close to $\tau$. For each $\tau$ of the plot, we filtered the two frequency traces with a band-pass filter centered on $\tau$. For a consistent correspondence between Allan deviation and correlation integration times, we chose the Allan deviation transfer function as the band-pass filter, defined as:

$$|H_A(f)|^2 = \frac{2 \sin^4 \pi\tau f}{(\pi\tau f)^2} \tag{3}$$

Finally, the correlation coefficient of the filtered frequency traces $f_1$ and $f_2$, each of length *N*, was defined by [56]:

$$corr_{f_1 f_2} = \frac{\sum_{i=1}^{N}(f_{1,i} - \bar{f}_1)(f_{2,i} - \bar{f}_2)}{N s_{f1} s_{f2}} \qquad (4)$$

Where $\bar{f}_1$ and $\bar{f}_2$ are the sample means of $f_1$ and $f_2$, respectively, and $s_{f1}$ and $s_{f2}$ are their standard deviations.

**REFERENCES**

1. Teufel, J. D. *et al.* Sideband cooling of micromechanical motion to the quantum ground state. *Nature* **475,** 359–63 (2011).

2. O'Connell, A. D. *et al.* Quantum ground state and single-phonon control of a mechanical resonator. *Nature* **464,** 697–703 (2010).

3. Lifshitz, R. & Cross, M. C. in *Nonlinear Dynamics of Nanosystems* 221–266 (Wiley-VCH Verlag GmbH & Co. KGaA, 2010).

4. Kacem, N. *et al.* Overcoming limitations of nanomechanical resonators with simultaneous resonances. *Appl. Phys. Lett.* **073105,** (2015).

5. Moser, J. *et al.* Ultrasensitive force detection with a nanotube mechanical resonator. *Nat. Nanotechnol.* **8,** 493–6 (2013).

6. Chaste, J. *et al.* A nanomechanical mass sensor with yoctogram resolution. *Nat. Nanotechnol.* **7,** 301–4 (2012).

7. Cleland, A. N. & Roukes, M. L. Noise Processes in Nanomechanical Resonators. *J. Appl. Phys.* **92,** 2758–2769 (2002).

8. Vig, J. R. & Kim, Y. Noise in Microelectromechanical System Resonators. *IEEE Trans. Ultrason. Ferroelectr. Freq. Control* **46,** 1558–1565 (1999).

9. Atalaya, J., Isacsson, A. & Dykman, M. I. Diffusion-induced dephasing in nanomechanical resonators. *Phys. Rev. B* **045419,** 1–9 (2011).

10. Malvar, O. *et al.* Tapered silicon nanowires for enhanced nanomechanical sensing. *Appl. Phys. Lett.* **033101,** (2013).

11. Bartsch, S. T., Rusu, A. & Ionescu, A. M. Phase-locked loop based on

nanoelectromechanical resonant-body field effect transistor. *Appl. Phys. Lett.* **101,** 153116 (2012).

12. Kumar, M. & Bhaskaran, H. Ultrasensitive Room-Temperature Piezoresistive Transduction in Graphene-Based Nanoelectromechanical Systems. *Nano Lett.* (2015). doi:10.1021/acs.nanolett.5b00129

13. Ekinci, K. L., Yang, Y. T. & Roukes, M. L. Ultimate limits to inertial mass sensing based upon nanoelectromechanical systems. *J. Appl. Phys.* **95,** (2004).

14. Robins, W. P. Phase Noise in Signal Sources. *Electronics and Power* **30,** 82 (1984).

15. Gouttenoire, V. *et al.* Digital and FM demodulation of a doubly clamped single-walled carbon-nanotube oscillator: towards a nanotube cell phone. *Small* **6,** 1060–5 (2010).

16. Mile, E. *et al.* In-plane nanoelectromechanical resonators based on silicon nanowire piezoresistive detection. *Nanotechnology* **21,** 165504 (2010).

17. Maizelis, Z. A., Roukes, M. L. & Dykman, M. I. Detecting and characterizing frequency fluctuations of vibrational modes. *Phys. Rev. B* **84,** 144301 (2011).

18. Zhang, Y., Moser, J., Güttinger, J., Bachtold, A. & Dykman, M. I. Interplay of Driving and Frequency Noise in the Spectra of Vibrational Systems. *Phys. Rev. Lett.* **113,** 255502 (2014).

19. Schneider, B. H., Singh, V., Venstra, W. J., Meerwaldt, H. B. & Steele, G. A. Observation of decoherence in a carbon nanotube mechanical resonator. *Nat. Commun.* 1–5 (2014). doi:10.1038/ncomms6819

20. Dykman, M. I., Khasin, M., Portman, J. & Shaw, S. W. Spectrum of an Oscillator with Jumping Frequency and the Interference of Partial Susceptibilities. *Phys. Rev. Lett.* **105,** 230601 (2010).

21. Yang, Y. T., Callegari, C., Feng, X. L. & Roukes, M. L. Surface adsorbate fluctuations and noise in nanoelectromechanical systems. *Nano Lett.* **11,** 1753–9 (2011).

22. Fong, K. Y., Pernice, W. H. P. & Tang, H. X. Frequency and phase noise of ultrahigh Q silicon nitride nanomechanical resonators. *Phys. Rev. B* **85,** 161410 (2012).

23. Siria, A. *et al.* Electron Fluctuation Induced Resonance Broadening in Nano Electromechanical Systems: The Origin of Shear Force in Vacuum. *Nano Lett.* **12,** 3551–3556 (2012).

24. Steele, G. A. *et al.* Strong Coupling Between Single-Electron Tunneling and Nanomechanical Motion. *Science (80-. ).* **325,** 1103–1108 (2009).

25. Miao, T., Yeom, S., Wang, P., Standley, B. & Bockrath, M. Graphene Nanoelectromechanical Systems as Stochastic-Frequency Oscillators. *Nano Lett.* **14,** 2982–2987 (2014).

26. Allan, D. W. Time And Frequency (Time-Domain) Characterization, Estimation, And Prediction Of Precision Clocks And Oscillators. *IEEE Trans. Ultrason. Ferroelectr. Freq. Control* **34,** 647–654 (1987).

27. Postma, H. W. C., Kozinsky, I., Husain, A. & Roukes, M. L. Dynamic Range of Nanotube and Nanowire-Based Electromechanical Systems. *Appl. Phys. Lett.* **86,** 1–3 (2005).

28. Jensen, K., Kim, K. & Zettl, A. An atomic-resolution nanomechanical mass sensor. *Nat. Nanotechnol.* **3,** 533–537 (2008).

29. Ramos, D. *et al.* Optomechanics with Silicon Nanowires by Harnessing Confined Electromagnetic Modes. *Nano Lett.* **12,** 932–937 (2012).

30. Olcum, S. *et al.* Weighing nanoparticles in solution at the attogram scale. *Proc. Natl. Acad. Sci.* (2014). doi:10.1073/pnas.1318602111

31. Ivaldi, P. *et al.* 50 nm thick AlN film-based piezoelectric cantilevers for gravimetric detection. *J. Micromechanics Microengineering* **21,** 085023 (2011).

32. Burg, T. P. *et al.* Weighing of biomolecules, single cells and single nanoparticles in fluid. *Nature* **446,** 1066–9 (2007).

33. Chiu, H., Hung, P., Postma, H. W. C. & Bockrath, M. Atomic-Scale Mass Sensing Using Carbon Nanotube Resonators. *Nano Lett.* **8,** 4342–4346 (2008).

34. Sansa, M., Fernández-Regúlez, M., Llobet, J., San Paulo, Á. & Pérez-Murano, F. High-sensitivity linear piezoresistive transduction for nanomechanical beam resonators. *Nat. Commun.* **5,** (2014).

35. Hanay, M. S. *et al.* Single-protein nanomechanical mass spectrometry in real time. *Nat. Nanotechnol.* **7,** 602–608 (2012).

36. Naik, A. K., Hanay, M. S., Hiebert, W. K., Feng, X. L. & Roukes, M. L. Towards single-molecule nanomechanical mass spectrometry. *Nat. Nanotechnol.* **4,** 445–50 (2009).

37. Feng, X. L., White, C. J., Hajimiri, a & Roukes, M. L. A self-sustaining ultrahigh-frequency nanoelectromechanical oscillator. *Nat. Nanotechnol.* **3,** 342–6 (2008).

38. Yang, Y. T., Callegari, C., Feng, X. L., Ekinci, K. L. & Roukes, M. L. Zeptogram-scale nanomechanical mass sensing. *Nano Lett.* **6,** 583–6 (2006).

39. Gray, J. M., Bertness, K. a., Sanford, N. a. & Rogers, C. T. Low-frequency noise in gallium nitride nanowire mechanical resonators. *Appl. Phys. Lett.* **101,** 233115 (2012).

40. Verd, J. *et al.* Monolithic CMOS MEMS Oscillator Circuit for Sensing in the Attogram Range. *Electron Device Lett. IEEE* **29,** 146–148 (2008).

41. Verd, J. *et al.* Design , Fabrication , and Characterization of a Submicroelectromechanical Resonator With Monolithically Integrated CMOS Readout Circuit. *J. Microelectromechanical Syst.* **14,** 508–519 (2005).

42. Larsen, T., Schmid, S., Villanueva, L. G. & Boisen, A. Photothermal Analysis of Individual Nanoparticulate Samples Using Micromechanical Resonators. *ACS Nano* **7,** 6188–6193 (2013).

43. Gavartin, E., Verlot, P. & Kippenberg, T. J. Stabilization of a linear nanomechanical oscillator to its thermodynamic limit. *Nat. Commun.* **4,** (2013).

44. Villanueva, L. G. *et al.* A nanoscale parametric feedback oscillator. *Nano Lett.* **11,** 5054–9 (2011).

45. Chen, C. *et al.* Performance of monolayer graphene nanomechanical resonators with electrical readout. *Nat. Nanotechnol.* **4,** 861–867 (2009).

46. Park, K. K. *et al.* Capacitive micromachined ultrasonic transducer (CMUT) as a chemical sensor for DMMP detection. *Sensors Actuators B Chem.* **160,** 1120–1127 (2011).

47. Sage, E. *et al.* Neutral particle mass spectrometry with nanomechanical systems. *Nat. Commun.* 1–5 (2015). doi:10.1038/ncomms7482

48. Moser, J., Eichler, A., Güttinger, J., Dykman, M. I. & Bachtold, A. Nanotube mechanical resonators with quality factors of up to 5 million. *Nat. Nanotechnol.* **9,**

1007–1011 (2014).

49. Vinante, A. Thermal frequency noise in micromechanical resonators due to nonlinear mode coupling. *Phys. Rev. B* **90,** 024308 (2014).

50. Venstra, W. J., van Leeuwen, R. & van der Zant, H. S. J. Strongly coupled modes in a weakly driven micromechanical resonator. *Appl. Phys. Lett.* **101,** 243111 (2012).

51. Barnard, A. W., Sazonova, V., van der Zande, A. M. & McEuen, P. L. Fluctuation broadening in carbon nanotube resonators. *Proc. Natl. Acad. Sci. U. S. A.* **109,** 19093–6 (2012).

52. Dykman, M. I. & Krivoglaz, M. A. Classical Theory of Nonlinear Oscillators Interacting with a Medium. *Phys. Status Solidi* **48,** 497 (1971).

53. Zhang, Y. & Dykman, M. I. Spectral effects of dispersive mode coupling in driven mesoscopic systems. *Phys. Rev. B* **92,** 1–16 (2015).

54. Yazdanian, S. M., Hoepker, N., Kuehn, S., Loring, R. F. & Marohn, J. A. Quantifying Electric Field Gradient fluctuations over polymers using Ultrasensitive Cantilevers. *Nano Lett.* **9,** 22732279 (2009).

55. Nichol, J. M., Hemesath, E. R., Lauhon, L. J. & Budakian, R. Nanomechanical detection of nuclear magnetic resonance using a silicon nanowire oscillator. *Phys. Rev. B* **85,** 1–6 (2012).

56. Papoulis, A. & Pillai, S. U. *Probability, Random Variables, and Stochastic Processes, Fourth Edition*. (2002).

**ACKNOWLEDGMENTS**

The authors are grateful for partial support from the LETI Carnot Institute NEMS-MS project, as well as from the European Union through the ERC-Enlightened project (616251) and the Marie-Curie Eurotalents outgoing (S.H.) and ingoing (M.S.) fellowships. They also thank Carine Marcoux and Cécilia Dupré for their support with the device fabrication. L.G.V. acknowledges financial support from the Swiss National Science Foundation (PP00P2-144695). A.K.N. acknowledges financial support from Indian Institute of Science, Bangalore.